\documentclass[conference,letterpaper]{IEEEtran}

\usepackage{cite}
\usepackage{booktabs}
\usepackage{graphicx}
\usepackage{amssymb}
\usepackage{amsfonts}
\usepackage{amsmath}
\usepackage{epsfig}
\usepackage{color}
\usepackage{fancybox}
\usepackage{multirow}
\usepackage{setspace}
\usepackage{psfrag}
\usepackage[ruled,vlined,linesnumbered]{algorithm2e}
\usepackage{hyperref}
\usepackage{mathrsfs}
\usepackage{subfigure}
\usepackage{graphicx}
\usepackage{bm}
\usepackage{algorithmic}

\ifCLASSINFOpdf

\else

\fi

\begin{document}
\title{
Simultaneous Localization and Mapping Using Active mmWave Sensing in 5G NR
}

\author{\IEEEauthorblockN{Tao Du\textsuperscript{1}, Jie Yang\textsuperscript{2,3}, Fan Liu\textsuperscript{1}, Jiaxiang Guo\textsuperscript{3}, Shuqiang Xia\textsuperscript{4,5}, Chao-Kai Wen\textsuperscript{6} and Shi Jin\textsuperscript{1,2}}
\IEEEauthorblockA{1. National Mobile Communications Research Laboratory, Southeast University, Nanjing, 210096, China\\
	2.\hspace{-0.6mm} Frontiers\hspace{-0.6mm} Science\hspace{-0.6mm} Center\hspace{-0.6mm} for\hspace{-0.6mm} Mobile\hspace{-0.6mm} Information\hspace{-0.6mm} Communication\hspace{-0.6mm} and\hspace{-0.6mm} Security,\hspace{-0.6mm} Southeast\hspace{-0.6mm} University,\hspace{-0.6mm} Nanjing\hspace{-0.6mm} 211189, \hspace{-0.6mm}China\\
	3.\hspace{-0.6mm} Key\hspace{-0.6mm} Laboratory\hspace{-0.6mm} of\hspace{-0.6mm} Measurement\hspace{-0.6mm} and\hspace{-0.6mm} Control\hspace{-0.6mm} of\hspace{-0.6mm} Complex\hspace{-0.6mm} Systems\hspace{-0.6mm} of\hspace{-0.6mm} Engineering,\hspace{-0.6mm} Southeast\hspace{-0.6mm} University, \hspace{-0.6mm}Nanjing\hspace{-0.6mm} 211189,\hspace{-0.6mm} China\\
	4. ZTE Corporation, Shenzhen 518057, China\\
	5. State Key Laboratory of Mobile Network and Mobile Multimedia, Shenzhen 518057, China\\
	6. Institute of Communications Engineering, National Sun Yat-sen University, Kaohsiung 80424, Taiwan\\
Email:\hspace{-1.5mm} \{dutao,\hspace{-1.5mm} yangjie,\hspace{-1.5mm} 220234853,\hspace{-1.5mm} jinshi\}\hspace{-0.5mm}@\hspace{-0.5mm}seu.edu.cn;\hspace{-1.5mm} fan.liu\hspace{-0.5mm}@\hspace{-0.5mm}seu.edu.cn;\hspace{-1.5mm} xia.shuqiang\hspace{-0.5mm}@\hspace{-0.5mm}zte.com.cn;\hspace{-1.5mm} chaokai.wen\hspace{-0.5mm}@\hspace{-0.5mm}mail.nsysu.edu.tw}
}

\maketitle

\begin{abstract}
Millimeter-wave (mmWave) 5G New Radio (NR) communication systems, with their high-resolution antenna arrays and extensive bandwidth, offer a transformative opportunity for high-throughput data transmission and advanced environmental sensing. Although passive sensing-based SLAM techniques can estimate user locations and environmental reflections simultaneously, their effectiveness is often constrained by assumptions of specular reflections and oversimplified map representations. To overcome these limitations, this work employs a mmWave 5G NR system for active sensing, enabling it to function similarly to a laser scanner for point cloud generation. Specifically, point clouds are extracted from the power delay profile estimated from each beam direction using a binary search approach. To ensure accuracy, hardware delays are calibrated with multiple predefined target points. Pose variations of the terminal are then estimated from point cloud data gathered along continuous trajectory viewpoints using point cloud registration algorithms. Loop closure detection and pose graph optimization are subsequently applied to refine the sensing results, achieving precise terminal localization and detailed radio map reconstruction. The system is implemented and validated through both simulations and experiments, confirming the effectiveness of the proposed approach.

\end{abstract}

\IEEEpeerreviewmaketitle

\section{Introduction}
The convergence of wireless communication and radar sensing is rapidly becoming a reality. Future wireless systems are expected to incorporate radar-like architectures in hardware, frequency, and signal processing, paving the way for enhanced sensing capabilities~\cite{lu2024integrated}. This integration, known as integrated sensing and communication (ISAC), promises significant improvements in spectrum and energy efficiency, which has spurred considerable research interest.

Radio simultaneous localization and mapping (SLAM) leverages communication network sensing for localizing mobile agents while mapping their surroundings~\cite{amjad2023radio}. The BP-SLAM algorithm in \cite{leitinger2019belief} uses random vectors to represent environmental features, while \cite{kim20205g} applies Random Finite Sets and the Probability Hypothesis Density filter for data association.

Hybrid strategies, such as in~\cite{yang2022hybrid}, improve passive sensing by incorporating active priors. The mapping performance and the accuracy of UE state estimation were improved by correlating the passive sensing features with the point cloud actively sensed by static base station (BS) in~\cite{ge2023integrated}. Moreover, \cite{rastorgueva2024millimeter,yang2024isac} discuss the design and testing of mmWave-based SLAM systems. However, these mostly focus on passive sensing, relying on specular reflections and simplifying sensed features as virtual anchors (mirror images of BSs). Although efficient for storage, this approach lacks detailed environmental shape information, complicating consistent mapping across multiple BSs.

Active sensing, on the other hand, involves a mobile sensing terminal that actively emits signals and receives echoes from the environment, enabling a more flexible and autonomous sensing capability independent of external devices. For example, \cite{guidi2018indoor} presents an adaptive threshold grid mapping method that requires no prior environmental knowledge, while \cite{barneto2020radio} demonstrates 28\,GHz indoor imaging by combining measurements from multiple viewpoints. However, these studies often assume known measurement locations, limiting their applicability for agents without prior spatial knowledge.

Building upon these insights, this paper designs and validates a scheme for simultaneous terminal localization and radio point cloud mapping using active sensing with a 5G mmWave prototype system. The proposed system comprises a 28\,GHz mmWave beam management setup based on the 5G NR frame structure, utilizing an unused antenna array for environmental sensing through echo reception. To ensure accuracy, system hardware delays are calibrated using predefined target points. We then utilize orthogonal frequency-division multiplexing (OFDM) symbols to estimate the power delay profile (PDP) in single-target scenarios and employed coherent accumulation of multiple OFDM data payload symbols with matched filtering (MF) to reduce PDP sidelobes and enhance weak target detection in multi-target environments. Point clouds for each viewpoint were generated by extracting peaks from the estimated PDP using a binary search. Point cloud data across continuous trajectories were fused using point cloud registration to estimate terminal pose. Finally, precise terminal localization and detailed radio point cloud imaging of the environment were achieved through loop closure detection and pose graph optimization (PGO).

\section{System Model and Hardware Implementation}
\subsection{Geometry Model}\label{sec2-a}

\begin{figure*}[htbp]
	\centering
	\subfigure[System model]{
		\includegraphics[width=5cm]{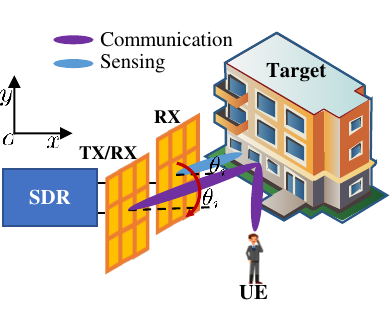}
		\label{fig1a}
	}
	\subfigure[Algorithmic processing flow]{
		\includegraphics[width=5cm]{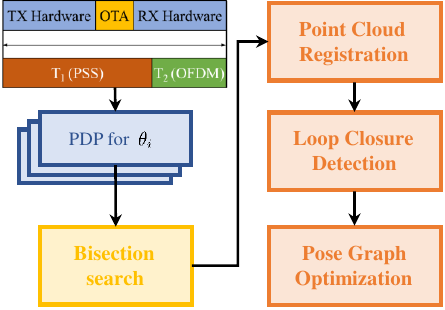}
		\label{fig1b}
	}
	\subfigure[Frame structure]{
		\includegraphics[width=6cm]{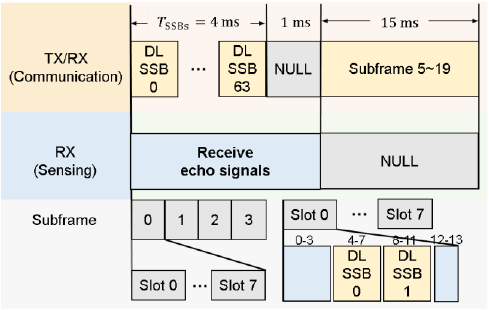}
		\label{fig1c}
	}
	\caption{System Architecture Design.}
	\label{fig1}
\end{figure*}

As illustrated in Fig.~\ref{fig1a}, the terminal includes two antenna arrays: TX/RX, operating in time-division duplex (TDD) mode for communication, and RX, which receives echoes from TX/RX transmissions reflected in the environment for sensing. In this active sensing model, mmWave signals transmitted by TX/RX reflect off a target and are received by RX. This transceiver setup, called the ISAC terminal, uses a radar-like approach, with both arrays sharing identical codebooks. For a target with index $k$, the round-trip time $\tau_k$ is derived from the received OFDM signal, allowing calculation of the target’s position. The target is located at an angle $\vartheta_{i}$ and a distance $c\tau_{k} /2$ from the terminal, where \(c\) is the speed of light. The TX/RX and RX arrays are separated to reduce self-interference, though increased spacing may lead to ranging errors, which are to be calibrated in future work.

\subsection{Channel Model}
In mmWave MIMO systems, spatial channel characteristics are described using steering vectors, representing the angular profile of plane waves. These vectors depend on the angles of arrival (AoA) and departure (AoD). Although our system uses a uniform planar array (UPA), beam scanning is confined to the horizontal plane, effectively making it behave like a uniform linear array (ULA). The steering vector for a normalized spatial angle \(\vartheta\) is defined as
\begin{equation}\label{eq2}
	\mathbf{a}(\vartheta) = \left[1, e^{-j2\pi \vartheta}, e^{-j4\pi \vartheta}, \ldots, e^{-j2\pi(N-1)\vartheta} \right]^\mathrm {T},
\end{equation}
where \({{\left( \cdot  \right)}^{\mathrm{T}}}\) indicates the transpose, \(N\) is the number of array elements, and \(\vartheta = \frac{d}{\lambda} \sin(\theta)\), with \(d\), \(\lambda\), and \(\theta\) as the antenna spacing, wavelength, and AoA, respectively. This expression models phase shifts due to different angles, crucial for array gain pattern formation.

The mmWave multipath channel is modeled by considering \(K\) distinct propagation paths. Each path is characterized by several parameters: AoA \(\theta_{\mathrm {R},k}\), AoD \(\theta_{\mathrm {T},k}\), delay \(\tau_k\), and complex gain \(b_k\). The frequency-domain mmWave multipath channel is then expressed as
\begin{equation}\label{eq4}
	\mathbf{H}(f) = \sum_{k=1}^{K} b_k e^{-j2\pi f \tau_k} \mathbf{a}_\mathrm {R}(\theta_{\mathrm {R},k}) \mathbf{a}_\mathrm {T}^\mathrm {H}(\theta_{\mathrm {T},k}),
\end{equation}
where \({{\left( \cdot  \right)}^{\mathrm{H}}}\) denotes the Hermitian transpose.

\subsection{Signal Model}
The transmitted signal \(\mathbf{x}_{i} \left ( t \right )\) propagates through multiple paths, resulting in the received signal \(\mathbf{y}_{i} \left ( t \right )\) expressed as
\begin{equation}
	\mathbf{y}_{i}(t) =  \sum_{k=1}^{K} b _k \mathbf{a}_\mathrm{R} (\theta_{\mathrm{R},k}) \mathbf{a}_\mathrm{T}^{\mathrm{H}}(\theta_{\mathrm{T},k}) \mathbf{x}_{i}(t - \tau_k) + \mathbf{z}(t),
\end{equation}
where \(\mathbf{z}(t)\) represents noise, and the steering vectors \(\mathbf{a}_\mathrm{R}\) and \(\mathbf{a}_\mathrm{T}\) define the spatial filtering at the receiver and transmitter, respectively. For the single radio frequency (RF) chain scenario considered, \(\mathbf{x}_{i}(t) = \mathbf{w}_{\mathrm{BF},i} x(t)\), where \(\mathbf{w}_{\mathrm{BF},i}\) is the analog beamforming vector of the \(i\)-th codebook, and \(x(t)\) is the single-stream transmitted signal. Then, the received signal after combiner is given by \(y_{i}(t) = \mathbf{w}_{\mathrm{BF},i}^{\mathrm{H}}\mathbf{y}_{i}(t)\).

The signal model for active range sensing with mmWave OFDM signals is structured as follows. The ISAC terminal uses phased arrays to steer its beam and capture reflections, enabling environment point cloud mapping based on the transmitted OFDM waveform from the TX/RX array. Signal processing leverages frequency-domain radar techniques, utilizing an ISAC frequency-domain resource grid with dimensions \(N_c \times R\), which holds the frequency-domain samples across \(N_c\) subcarriers and \(R\) OFDM symbols. On the TX/RX side, the baseband uplink waveform is generated by applying an inverse fast Fourier transform (IFFT) on the resource grid, followed by the addition of a cyclic prefix (CP) to the transmitted signal.

The RX uses a combiner aligned with the TX beamforming direction. The received signal \(y_{i}(t)\) is then demodulated and processed using FFT to extract the reflected signal in the frequency domain. The demodulated signal grid on the \(p\)-th subcarrier and \(q\)-th OFDM symbol is modeled as
\begin{equation}
	y_{i}^{\left ( p,q \right ) } =  \sum_{k=1}^{K}  \varpi_{i,k} b_k e^{-j 2\pi p\tau_k \Delta f} \cdot x_{p,q} + z_{p,q},
\end{equation}
where \(x_{p,q}\) is the transmitted signal, \(z_{p,q}\) is the noise component, \(\Delta f\) denotes the subcarrier spacing and \(\varpi_{i,k} = \mathbf{w}_{\mathrm{BF},i}^{\mathrm{H}} \mathbf{a}_\mathrm{R} (\theta_{\mathrm{R},k}) \mathbf{a}_\mathrm{T}^{\mathrm{H}}(\theta_{\mathrm{T},k}) \mathbf{w}_{\mathrm{BF},i}  \) represents the degree of alignment between the transmitted beam direction and the target located at \(\theta_{k}\). A higher magnitude of \(\varpi_{i,k}\) indicate closer proximity to the target's actual azimuth. Then, the frequency-domain channel estimation \(\hat{\mathbf{h}}_{i}\) can be obtained from \( \{ y_{i}^{\left ( p,q \right ) }, \, p = 1, \dots, N_c  \} \) through channel estimation techniques.

\subsection{Hardware Implementation and Frame Structures}
To validate the feasibility of active sensing schemes, the experimental test was conducted over-the-air (OTA). The hardware architecture of the prototype system primarily consists of the SDR USRP 2974, the mmWave UPA mmPSA-TR64MX, and the Sync Node. Detailed specifications of the relevant hardware can be found in~\cite{yang2024isac}.

As mentioned in Section~\ref{sec2-a}, the TX/RX performs standard communication functions. During beam scanning, the dedicated sensing array, RX, utilizes an identical codebook to that of the TX/RX to receive echo signals, encompassing both synchronization signal blocks (SSBs) and OFDM symbols for sensing. Both the TX/RX and RX arrays are configured as \(8\times 8\) UPA. However, since only horizontal plane beam scanning is implemented, the effective aperture is equivalent to a ULA with \(N = 8\) elements. Beamforming is achieved using a 6-bit DFT codebook, resulting in the frame structure shown in Fig.~\ref{fig1c}. Each frame lasts 20 ms and consists of 20 subframes, with each subframe further divided into 8 time slots, and each slot containing 14 OFDM symbols. The hardware parameters are summarized in Table~\ref{tab2}.

To mitigate interference with existing 5G NR communication systems~\cite{yang2024isac}, sensing OFDM symbols are inserted within the reserved symbols allocated to the first 5 ms of each frame, as depicted by the light blue region in Fig.~\ref{fig1c}. Given that each radio frame contains 304 reserved OFDM symbols, and 47 codebook entries are selected for sensing in each antenna array orientation (detailed in Section~\ref{sec4-a}), each codebook entry is guaranteed 6 sensing symbols per frame. In our system, data from two consecutive frames are acquired to obtain 12 symbols per codebook entry for measurement.

\begin{table}[htbp!]
	\centering
	\caption{Hardware parameters for the mmWave prototype system}
	\label{tab2}
	\setlength\tabcolsep{4.8pt}
	\setlength{\tabcolsep}{6pt}
	\renewcommand{\arraystretch}{0.6}
	\footnotesize
	\begin{tabular}{cccc}
		\toprule
		Array size & \(8\times 8\) & Modulation & OFDM \\ \midrule
		Mapping  & 16\,QAM & RF & 28\,GHz \\ \midrule
		FFT size  & 1024 & Active subcarriers & 792 \\ \midrule
		
		Bandwidth& 95.04\,MHz & Subcarrier spacing & 120\,KHz   \\ \midrule
		
		Sampling rate  & 122.88\,MSps & Frame Duration & 20\,ms
		\\ \bottomrule
	\end{tabular}
\end{table}

\section{Radio Point Cloud Imaging}
\begin{figure}[htbp]
	\centering
	\resizebox{\linewidth}{!}{%
		\includegraphics{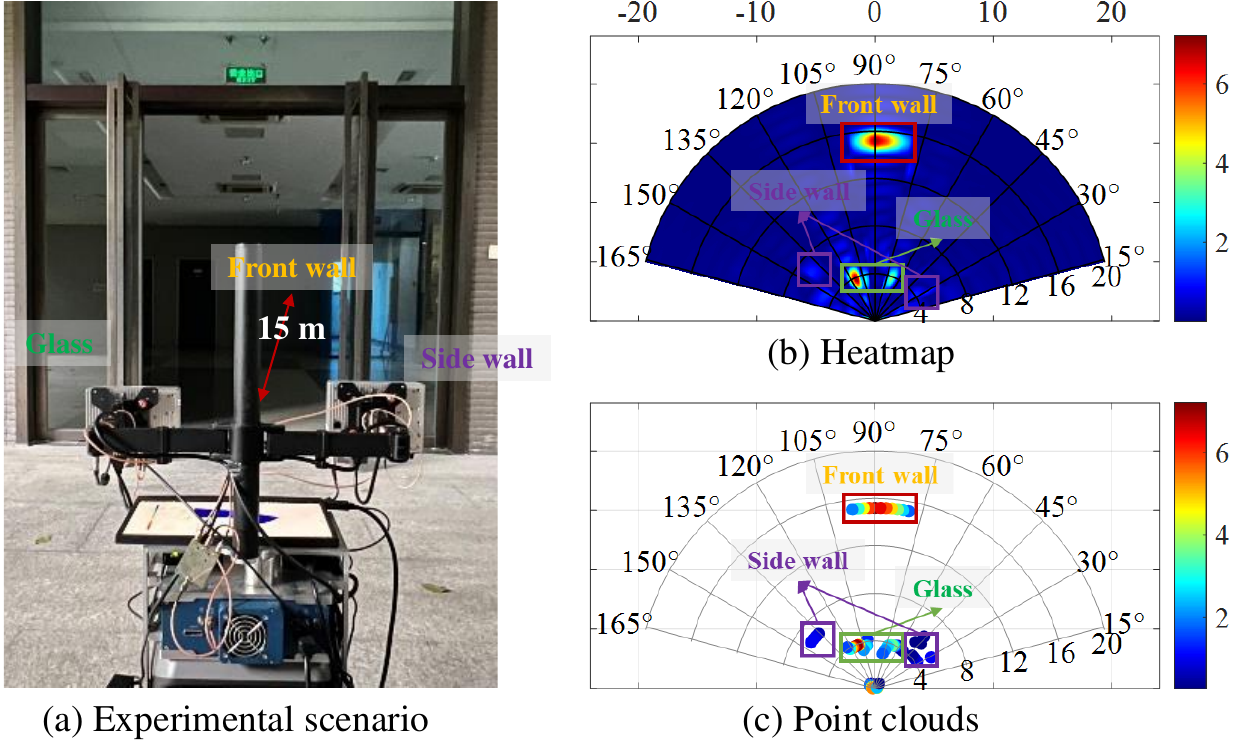} }%
	\caption{Experimental results of radio point cloud for a glass door.}
	\label{fig4}
\end{figure}

In this section, we introduce our scheme for obtaining radio point cloud imaging. As shown in Fig.~\ref{fig1a}, both the TX/RX and RX arrays of the ISAC terminal utilize a single RF chain. The TX and RX arrays are slightly offset spatially and employ a DFT codebook for analog beamforming. For a ULA with \(N\) elements, the \(M\)-bit DFT codebook is represented as \(\mathbf{W}_{\mathrm{BF}} =  \left [ \mathbf{w}_{\mathrm{BF},0},\dots ,\mathbf{w}_{\mathrm{BF},2^M-1} \right ]\), where \(\mathbf{w}_{\mathrm{BF},i} = \mathbf{a}(\vartheta_{i})/ \sqrt{N} \) with \(\mathbf{a}(\cdot )\) representing the steering vector defined in (\ref{eq2}) and \(\vartheta_{i} = i/2^M\) denoting the spatial angle corresponding to the \(i\)-th codebook. Using this discrete set of DFT codebooks, the array’s beam is steered toward various spatial directions to detect targets within the field of view (FoV). The system operates in full-duplex mode, mitigating self-interference through beam-space isolation. Sensing results from these varied directions are aggregated to create a radio heatmap image, or point cloud, as depicted in Fig~\ref{fig4}.

\subsection{Range Estimation Method for Single Target}
In scenarios with a single target per beam direction, the target range can be estimated by identifying the delay of the strongest path in the corresponding reflected signal. Within the wireless frame structure, an SSB is transmitted for each beam direction, primarily for OTA synchronization and beam indexing. Synchronization is achieved using a primary synchronization signal (PSS), which is a 127-point Zadoff-Chu (ZC) sequence.

Since ZC sequences have favorable auto- and cross-correlation properties, making them ideal for synchronization. The cross-correlation peak between the received signal and the locally stored ZC sequence marks the position of the synchronization header. Using the system’s sampling rate, the time delay of the reflected signal can then be calculated. This approach, however, introduces a grid error due to discrete sampling points, which is equal to \(c/\left ( 2 f_s \right )  = 1.22\)\,m.

To mitigate these grid errors, we can leverage the frequency-domain channel response of the OFDM symbols. Applying an IFFT to \(\hat{\mathbf{h}}_{i}\) yields the PDP in the time domain as \(h_{i}(t)=\sum_{k=1}^{K} \varpi_{i,k} b_{k} \delta \left ( t-\tau_{k} \right )\),
%\begin{equation}
%	h(t)=\sum_{k=1}^{K} b_{k} \delta \left ( t-\tau_{k} \right ),
%\end{equation}
where \(\delta\) represents the Dirac delta function. The PDP corresponding to the \(i\)-th codebook is denoted by \(\mathbf{p}_{i}\). Concatenating all \(\mathbf{p}_{i}\) in polar coordinates yields Fig.~\ref{fig4}(b), where the physical angle associated with each \(\mathbf{p}_{i}\) corresponds to the angle of maximum gain in its directional power distribution.

Combining the maximum values of each \(\mathbf{p}_{i}\) yields Fig.~\ref{fig4}(c), which is applicable only to single-target scenarios. For multi-target scenarios, successive cancellation can be employed to mitigate duplicate detections~\cite{yang2023angle}. The precision of delay estimation from the PDP depends on the signal-to-noise ratio (SNR), helping to reduce grid errors. However, due to limited frequency-domain bandwidth, the time-domain response appears as a sinc function rather than an ideal \(\delta\) function. The slower decay of the sinc function's sidelobes results in a range resolution of \(c/2(N_{c}\cdot \bigtriangleup f) = 1.22\)\,m, where, in the designed prototype system, \(N_{c} = 1024\) and \(\bigtriangleup f = 120\,\mathrm {KHz}\) as the number of OFDM subcarriers and subcarrier spacing in the prototype system, respectively.

While the PDP theoretically reduces grid errors, achieving this requires an infinitely long IFFT. Increasing the IFFT length, however, raises computational complexity and introduces greater overhead in locating the PDP peak. Here, we employ a fractional IFFT combined with a binary search to efficiently determine the PDP peak. Initially, we perform an \(N_c\) IFFT on \(\mathbf{H}\) to obtain discrete PDP samples, extracting the \(\hat{K}\) largest peaks. For each peak, we identify the adjacent point with the next highest magnitude and compute a fractional IFFT at the midpoint between the peak and this adjacent point. If the magnitude at this midpoint surpasses that of the current peak, it replaces the peak; otherwise, it replaces the adjacent point. This iterative process continues until the desired precision or a specified number of iterations is reached.

\subsection{Range Estimation Method for Multiple Targets}
\begin{figure}[t]
	\centering
	\resizebox{2.7in}{!}{%
		\includegraphics{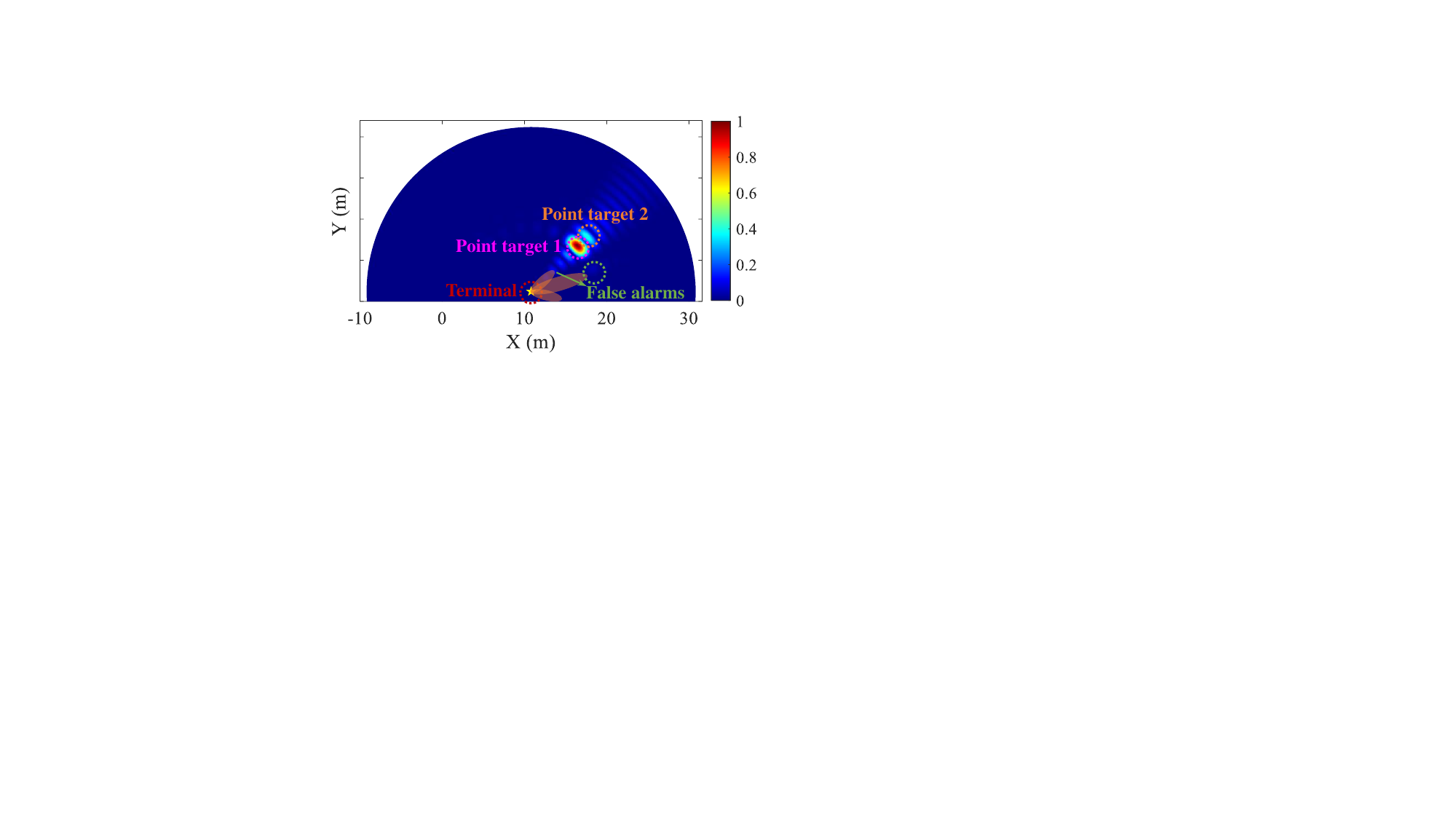} }%
	\caption{The impact of PDP sidelobes and beam sidelobes on target detection.}
	\label{fig8}
\end{figure}

In multi-target scenarios, the slow decay of the sinc function's sidelobes can obscure weaker targets, as the sidelobes of stronger targets may overlap with the main peaks of weaker targets, leading to missed detections. For instance, as illustrated in Fig.~\ref{fig8}, two point targets are positioned at 8\,m and 9\,m along the 45$^\circ$ direction, with respective path gains of 1 and 0.3, respectively. The PDP of the target at 9\,m is substantially obscured by that of the target at 8\,m, resulting in an apparent shift of the highlighted region toward the 10\,m position. MF can enhance detection performance in such scenarios, as it maximizes the SNR based on known signal characteristics. This approach is particularly effective in low-SNR environments, making it suitable for detecting weaker targets in the presence of stronger ones.

For brevity, the beam index will be omitted in the subsequent discussion. Considering a scenario where the received signal comprises \(K\) target echo components, it can be represented in the discrete-time domain as
\begin{equation}
	\mathbf{y} =   \sum_{k=1}^{K} b_{k}\mathbf{J}_{\widetilde{\tau}_{k}}\widetilde{\mathbf{x}} + \mathbf{z},
\end{equation}
where \(\widetilde{\mathbf{x}}\) contains \(N_c\) time-domain samples of a single OFDM symbol, \(\mathbf{z}\) is the additive white Gaussian noise, \(\widetilde{\tau}_{k}\) denotes the time delay of the \(k\)-th target, normalized with respect to the sampling interval, and \(\mathbf{J}_{k}\) is the circular shift matrix of \(k\) samples~\cite{stoica2009designing}. The inclusion of the CP results in a circular shift of \(\widetilde{\tau}_{k}\) samples for the received signal component corresponding to the target with delay \(\tau_{k}\). Accordingly, the \(i\)-th MF output may be obtained by muliplying \(\mathbf{y}\) with a delayed version of \(\widetilde{\mathbf{x}}\), expressed as
\begin{align}
	\widetilde{r}_{i} &\nonumber = \widetilde{\mathbf{x}}^{\mathrm{H}}\mathbf{J} _{i}^{\mathrm{T}}\mathbf{y} =   \sum_{k=1}^{K} b_{k}\widetilde{\mathbf{x}}^{\mathrm{H}}\mathbf{J}_{\widetilde{\tau}_{k}-i}\widetilde{\mathbf{x}} + \widetilde{\mathbf{x}}^{\mathrm{H}}\mathbf{J} _{i}^{\mathrm{T}}\mathbf{z}\\
	& \triangleq  \sum_{k=1}^{K} b_{k}G_{\widetilde{\tau}_{k}-i} + \widetilde{z}_i, \quad \forall i,
\end{align}
where \(G_i = \widetilde{\mathbf{x}}^{\mathrm{H}}\mathbf{J}_{i}\widetilde{\mathbf{x}}\) is the periodic auto-correlation function of \(\widetilde{\mathbf{x}}\), and \(\widetilde{z}_i\) is the output noise. The targets may be detected by searching over the range profile \(|\widetilde{r}_i|^2\), which is expected to generate high peaks at \(i = \widetilde{\tau}_{k}\), and low sidelobes elsewhere.

Since \(R\) consecutive OFDM symbols are transmitted in each direction, we may further improve the MF performance by coherently integrating \(R\) MF outputs. Thanks to the independence of \(R\) symbols, such coherent intergation considerably suppress the sidelobe level incurred by random data symbols, as well as the noise level. This enhances the weak target detection performance, particularly in low-SNR scenarios.

\subsection{Hardware Delay Bias Estimation}
Although the designed prototype system operates in a monostatic configuration, the delay estimation includes a fixed bias introduced by the hardware. As shown in Fig.~\ref{fig1b}, the estimated delay from the received signal comprises two components: a hardware-induced delay bias and an OTA delay, the latter being directly related to the target range.

To estimate the fixed hardware delay, targets were positioned at various distances, and the difference between the measured and theoretical time delays was analyzed. Both the PSS and OFDM symbols are utilized to estimate the time delay. The PSS determines a coarse synchronization header position, \(T_1\), while the OFDM symbols refine this estimate by determining an offset, \(T_2\), within the PSS synchronization grid.

For synchronization, the received signal was downsampled by a factor of 4 before determining the PSS synchronization header position. Although the PSS occupies only 127 subcarriers, the downsampled rate still satisfies the Nyquist sampling theorem. This approach allows for rapid synchronization while ensuring that the starting sample of the received signal remains within the cyclic prefix duration. Given \(N_{\mathrm{T}}\) target measurements, the estimated hardware delay \(T_{\mathrm {H}}\) is calculated as
\begin{equation}\label{eq11}
	T_{\mathrm {H}} = \frac{1}{N_{\mathrm{T}}}  \sum_{k=1}^{N_{\mathrm{T}}}  \left ( T_1^{(i)} + T_2^{(i)} - T_{\mathrm {O}}^{(i)} \right ),
\end{equation}
where \(T_{\mathrm {O}}^{(i)}\), \(T_1^{(i)}\) and \(T_2^{(i)}\) represent the OTA time delay, the PSS delay estimate, and the OFDM delay estimate for the \(i\)-th target, respectively.

\section{Radio Point Cloud Imaging based SLAM}\label{sec4}
This section describes the generation and processing of the radio point cloud for SLAM. We begin by detailing how the radio point cloud is generated and processed. Then, a laser point cloud SLAM framework is employed to fuse multi-point measurements across consecutive trajectories, allowing for the estimation of the terminal's trajectory and the reconstruction of the environment map.

\subsection{Radio Point Cloud Preprocessing}\label{sec4-a}
Since the UPA covers only the region facing the array aperture, achieving a full 360$^\circ$ point cloud at a single location requires rotating the ISAC terminal to different orientations, performing measurements at each orientation, and stitching the acquired point clouds together. The beams generated by the DFT codebook become wider and produce higher sidelobes as the steering direction approaches parallel to the array plane. To minimize the impact of these sidelobes on measurements, 23 beams on either side of the array normal direction are selected for each array orientation, covering a range of $\pm 45.95^\circ$ relative to the array normal. Measurements are taken with the array oriented at \(0\), \(\pi/2\), \(\pi\) and \(3\pi/2\), providing full spatial coverage.

For channel modeling, we employ the standard far-field multipath channel model in (\ref{eq4}). Similar to most ray-tracing software, channel parameters such as angles and delays are derived by casting a dense set of rays. In this simulation, a total of 188 rays (\(47\times 4\)) are used. Each ray is parameterized by \(\theta_{k}\) and \(\tau_k\), representing the steering angle associated with the \(k\)-th DFT codebook entry and the time delay of the closest target at that angle, respectively.

Once measurements are collected from all four orientations, producing a comprehensive spatial PDP distribution, we first normalize the PDP by its peak energy. Then, assuming a single target per direction, the target location is identified as the PDP peak, efficiently located using a bisection search. This process yields an initial point cloud for the given location. However, due to sidelobes, some points in the cloud may represent false detections, as illustrated by the green circle in Fig.~\ref{fig8}. Simulations using a 3D model in Blender, imported into the Sionna ray-tracing software~\cite{Sionna2022}, show that spurious points originating from sidelobes generally exhibit normalized energies below -13 dB. Points with normalized energies below this threshold are therefore discarded to refine the point cloud.

\subsection{Radio SLAM}

After preprocessing the point cloud, we obtain the full spatial distribution across consecutive trajectory points. A lidar point cloud SLAM algorithm is then employed to estimate the ISAC terminal's trajectory and reconstruct the surrounding wireless environment. Lidar SLAM systems typically comprise two main modules: the front-end and the back-end.

The front-end processes sequential scan data, estimating the relative pose transformations between consecutive time steps through a process known as scan matching. This module takes raw point cloud data as input and outputs the relative pose estimates between successive frames. Some implementations of the front-end also create local maps to aid subsequent scan matching and loop closure detection.

The back-end integrates the relative pose estimates from the front-end into a global map and trajectory. It maintains a pose graph where nodes represent the terminal’s poses and edges represent the relative transformations between these poses, derived from the front-end output and loop closure detection.

In this work, we implemented the lidar point cloud SLAM algorithm using MATLAB's Robotics System Toolbox. The system represents the environment with a grid map and employs scan matching and pose graph optimization techniques to enhance accuracy and robustness. As the terminal moves and acquires new mmWave beam scan data, a scan matching algorithm aligns the current scan with previous scans or pre-built submaps. This scan matching typically follows a two-stage approach: an initial coarse pose estimate is obtained through a computationally efficient grid-based matching method, followed by a refinement stage.

For efficient loop closure detection, the system uses a submap strategy, where a fixed number of consecutive scans are aggregated into a submap, reducing the number of comparisons needed during loop closure detection. Loop closures are identified by comparing the similarity between the current scan and previous submaps. Finally, a graph optimization algorithm refines the pose graph, minimizing errors in pose constraints and yielding a globally consistent trajectory and map. This pose graph optimization (PGO) effectively reduces accumulated errors, enhancing the accuracy and robustness of the SLAM system.
\begin{figure}[htbp]
	\centering
	\resizebox{3in}{!}{%
		\includegraphics{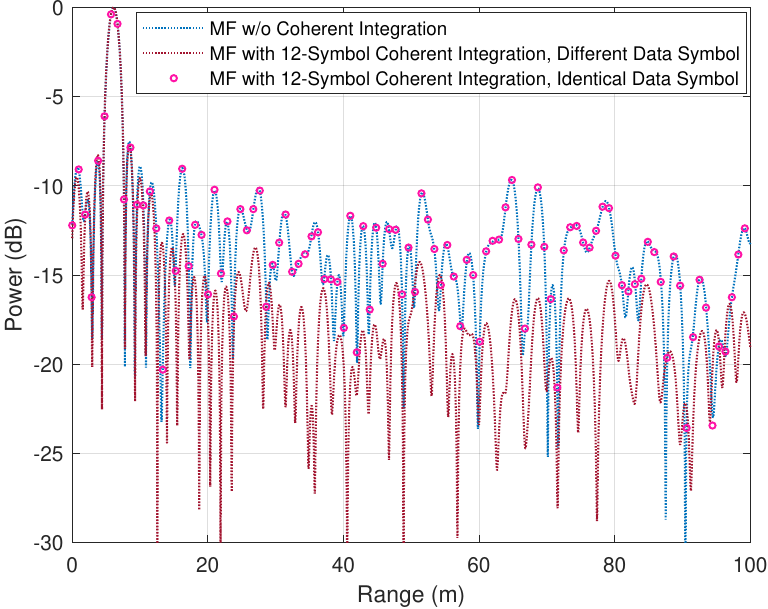} }%
	\caption{Comparison of MF sidelobe levels with experimental results}
	\label{fig5}
\end{figure}

\begin{figure*}[htbp]
	\centering
	\resizebox{6.2in}{!}{
		\includegraphics{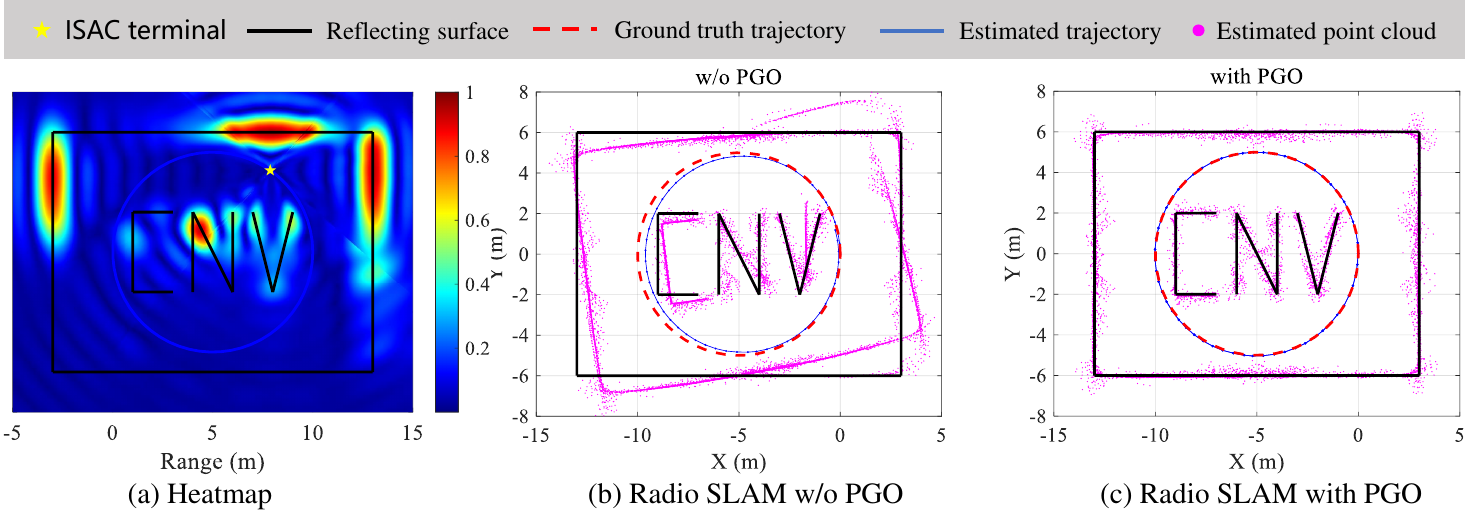} }
	\caption{Simulation results of radio point cloud imaging-based SLAM.}
	\label{fig6}
\end{figure*}

\section{Experiments and Simulations}
\subsection{Experimental Setup and Results}
To obtain radio beam scan data at individual locations, we developed a monostatic mmWave radar scanning system. Initially, the system's hardware delay was estimated to be approximately \(32\,\mu s\) using 16 target points spaced at 0.8m intervals, calculated according to (\ref{eq11}). After compensating for this delay, the ranging errors using $N_c$-point IFFT (\(\varepsilon_{1}\)) and bisection search (\(\varepsilon_{2}\)) are summarized in Table~\ref{tab1}. The measurement procedure involved positioning a \(1\, {\rm m}^2\) metal plate along the array face’s normal direction and incrementally shifting it by 0.8\,m per step. The estimated distance \(\hat{d}\) for the normal beam was extracted from the point cloud, and the difference between \(\hat{d}\) and the ground truth provided the estimation errors \(\varepsilon_{1}\) and \(\varepsilon_{2}\). Results indicate that the bisection search significantly enhances estimation accuracy, achieving sub-0.5\,m precision for single-target estimation with a RMSE of 0.34\,m.
\begin{table}[htbp!]
	\centering
	\caption{Ranging Error for a Single Metal Plate}
	\label{tab1}
	\setlength\tabcolsep{4.8pt}
	\setlength{\tabcolsep}{4pt}
	\renewcommand{\arraystretch}{0.5}
	\footnotesize
	\begin{tabular}{ccccccccc}
		\toprule
		Range (m) & 0.8 & 1.6 & 2.4 & 3.2 & 4.0 & 4.8 & 5.6 & 6.4  \\ \midrule
		\(\varepsilon_{1}\)  & -0.94 & 0.70 & \textbf{0.10} & 0.32 & -0.48 & \textbf{-0.06} & -0.86 & 0.78  \\ \midrule
		
		\(\varepsilon_{2}\)  & \textbf{-0.54} & \textbf{0.41} & 0.24 & \textbf{0.03} & \textbf{-0.13} & -0.31 & \textbf{-0.48} & \textbf{0.43} \\ \midrule

		Range (m) & 7.2 & 8.0 & 8.8 & 9.6 & 10.4 & 11.2 & 12.0 & 12.8  \\ \midrule
		\(\varepsilon_{1}\) & \textbf{-0.02} & 0.40 & -0.40 & \textbf{0.02} & -0.78 & 0.86 & \textbf{0.06} & 0.49   \\ \midrule
		
		\(\varepsilon_{2}\) & 0.27 & \textbf{0.07} & \textbf{-0.10} & -0.27 & \textbf{-0.46} & \textbf{-0.46} & 0.27 & \textbf{0.11}
		\\ \bottomrule
	\end{tabular}
\end{table}
Experiments were then conducted in a room featuring a glass door, as depicted in Fig.~\ref{fig4}. The glass door, positioned along the normal of the antenna array, was open during testing. A concrete wall was located 15\,m from the ISAC terminal within the room, with additional side glass and concrete walls extending from either side of the glass door. As shown in the central radar heatmap, the system accurately detected both the wall and glass door. A radio point cloud map, displayed on the right, was generated by extracting the peak PDP values across various directions in the heatmap. However, the point cloud exhibited some distortion at large angular deviations from the array's normal, primarily due to weaker reflected energy and sidelobe interference at off-normal angles, as discussed in Section \ref{sec4-a}.

Fig.~\ref{fig5} compares results obtained using a single OFDM symbol with those obtained through the coherent integration of 12 OFDM symbols. The data OFDM symbols are generated randomly, where ``\emph{different}'' denotes distinct data across symbols and ``\emph{identical}'' denotes consistent data across symbols. The MF approach of different data symbols effectively reduces sidelobe levels by approximately 5 dB, particularly enhancing performance at lower SNR. The superior performance observed with different data, compared to identical data, stems from the reduction in sidelobes achieved through the accumulation of randomized data. These findings demonstrate the feasibility of employing analog beamforming for radar-based active sensing and point cloud map generation, with SNR-dependent improvements facilitated by the MF technique.

\subsection{Simulation Setup and Results}
The simulated environment was designed to generate radio point cloud data along continuous terminal trajectories. The simulation parameters matched those of the mmWave prototype system, with a SNR of 10 dB. The simulated environment includes surrounding walls and central obstacles. The ISAC terminal follows a circular trajectory of 60 equidistant points with a radius of 5 meters. As shown in Fig.~\ref{fig6}(a), high reflection energy is observed at normal incidence with walls, while reflection energy decreases at off-normal angles due to reduced sidelobe gain.

By fusing the radio point clouds from consecutive positions using a lidar point cloud SLAM algorithm, the results in Figs.~\ref{fig6}(b) and \ref{fig6}(c) were obtained. The point cloud without PGO showed notable distortion and misalignment, failing to accurately depict the environmental structure, especially in peripheral areas where significant drift was evident. This misalignment reflects accumulated errors in trajectory or point cloud data. In contrast, the PGO-optimized point cloud showed accurate alignment with minimal distortion, preserving the environmental shape and clearly resolving the central ``CNV'' structure.

\section{Conclusion}
This paper presents a monostatic radar-based active sensing SLAM scheme for high-precision environmental mapping and localization. For single-target scenarios, we developed a joint PSS and OFDM delay estimation method with binary search for enhanced accuracy, while for multi-target scenarios, we applied a MF approach to reduce sidelobe interference.
A 3D point cloud map of the wireless environment was constructed using a DFT codebook and controlled antenna array rotation. By integrating data from a continuous trajectory and employing a laser SLAM algorithm for data fusion and optimization, we achieved accurate global trajectory and environmental mapping. Experimental validation using the mmWave prototype system demonstrated an RMSE of 0.34\,m, confirming the high accuracy and robustness of the proposed approach.

\section*{Acknowledgment}
This research was supported in part by the National Key Research and Development Program of China under Grant 2024YFE0200103, in part by the Fundamental Research Funds for the Central Universities 2242022k60004, in part by the National Natural Science Foundation of China (NSFC) under Grant 62301156 and 62331023, in part by the Key Technologies R\&D Program of Jiangsu (Prospective and Key Technologies for Industry) under Grants BE2023022 and BE2023022-1, and in part by the Guangdong Basic and Applied Basic Research Foundation under Grant 2024A1515011218. The work of C.-K. Wen was supported in part by the National Science and Technology Council of Taiwan through grant MOST 111-2221-E-110-020-MY3.

%\bibliographystyle{IEEEtran}
%\bibliography{ref}

% that's all folks
\end{document}